\documentstyle{article}

\newcommand{\BEQ}{\begin{equation}}
\newcommand{\EEQ}{\end{equation}}
\newcommand{\BEA}{\begin{eqnarray}}
\newcommand{\EEA}{\end{eqnarray}}
\newcommand{\BEAast}{\begin{eqnarray*}}
\newcommand{\EEAast}{\end{eqnarray*}}
\newcommand{\EA}{\end{array}}
\newcommand{\BA}{\begin{array}}
\newcommand{\Bb}{{\bf R}}
\newcommand{\Bcc}{{\bf C}}
\newcommand{\RZ}{\Bb^{2}}

\newcommand{\calG}{{\cal G}}
\newcommand{\calH}{{\cal H}}

\newcommand{\calF}{{\cal F}}
\newcommand{\einsi}{\frac{1}{i}}
\newcommand{\halb}{\frac{1}{2}}
\newcommand{\sgn}{{\rm sgn}}
\newcommand{\Hom}{{\rm Hom}}
\newcommand{\calFtild}{\tilde{\calF}}
\newcommand{\boldp}{{\mbox{\boldmath $p$}}}
\newcommand{\sboldp}{{\mbox{\boldmath ${\scriptstyle p}$}}}
\newcommand{\bfp}{{\mathchoice{\boldp}{\boldp}{\sboldp}{\sboldp}}}
\newcommand{\bft}{{\mbox{\boldmath $t$}}}
\newcommand{\ptild}{\tilde{\bfp}}
\newcommand{\pttild}{(\widetilde{\bfp,\bft})}
\newcommand{\pr}{{\rm pr}}
\newcommand{\prtild}{\tilde{\pr}}
\newcommand{\prttild}{\tilde{\tilde{\pr}}}
\newcommand{\xitild}{\tilde{\xi}}
\newcommand{\chitild}{\tilde{\chi}}
\newcommand{\calGtild}{\tilde{\calG}}
\newcommand{\Gtild}{\tilde{G}}
\newcommand{\MnDn}{M^{\times n}\setminus D_n}
\newcommand{\nMtild}{{}^n\tilde{M}}
\newcommand{\Fi}{\varphi}
\newcommand{\Hyp}{V^{+,m}}
\newcommand{\nRtildZ}{{}^n\tilde{\Bb}^2}
\newcommand{\calP}{{\cal P}}
\newcommand{\calM}{{\cal M}}
\newcommand{\Lor}{L_3^{\uparrow}}
\newcommand{\Lortild}{\tilde{L}_3^{\uparrow}}
\newcommand{\HypnTn}{(\Hyp)^{\times n}\:\times\:T^{\times n}}
\newcommand{\frakg}{{\mbox{\boldmath $g$}}}
\newcommand{\fraka}{{\mbox{\boldmath $a$}}}
\newcommand{\Ztilde}[1]{\stackrel{\scriptscriptstyle \approx}{#1}}
\newcommand{\rhobos}{\varrho_{\rm bos}}
\newcommand{\rhofer}{\varrho_{\rm fer}}
\newcommand{\Sbos}{S_{\rm bos}}
\newcommand{\Sfer}{S_{\rm fer}}
\newcommand{\utilde}[1]{\mathop{#1}\limits_{\widetilde{\phantom{\textstyle
 #1}}}}
\newcommand{\unity}{{\setlength{\unitlength}{1em}
                     \begin{picture}(0.75,1)
                     \put(0,0){$1$}
                     \put(0.34,0){\line(0,1){0.65}}
                     \end{picture}
                   }}
\newcommand{\semiprod}{\rhd\hspace*{-0.5em}<}
\renewcommand{\theequation}{\thesection.\arabic{equation}}
\newenvironment{Thm}%
{\medskip\par\noindent\refstepcounter{equation}{\bf Theorem (\theequation)}
\rm}%
{\par\medskip}
\newenvironment{Lem}{\medskip\par\noindent\refstepcounter{equation}{\bf Lemma
(\theequation)} \rm}{\par\medskip}
\newenvironment{Cor}{\medskip\par\noindent\refstepcounter{equation}{\bf
Corollary (\theequation)} \rm}{\par\medskip}
\newenvironment{Post}{\medskip\par\noindent\refstepcounter{equation}{\bf
Postulate (\theequation)} \rm}{\par\medskip}
\newenvironment{Ex}{\medskip\par\noindent\refstepcounter{equation}{\bf Example
(\theequation)} \rm}{\par\medskip}
\newenvironment{Proof}{\par\noindent {\bf Proof } \rm}{\par\medskip}

\begin{document}
\title{Hilbert Spaces for Nonrelativistic and Relativistic ``Free''
 Plektons (Particles with Braid Group Statistics)\thanks{To
  appear in the Proceedings of the Conference ``Advances in Dynamical Systems
  and Quantum Physics'', Capri (Italy) May 19-22 (1993)} }
\author{ J.~Mund\thanks{Supported by DFG, Sfb 288 ``Differentialgeometrie
  und Quantenphysik'' and by Studienstiftung des Deutschen Volkes} \\
  R.~Schrader\thanks{e-mail: schrader@vax1.physik.fu-berlin.dbp.de}\\
  Institut f\"{u}r theoretische Physik\\
  Freie Universit\"{a}t Berlin\\
  Germany}
\date{Dedicated to Gianfausto Dell'Antonio on the occasion\\
 of his 60{\em th} birthday}
\maketitle
\begin{abstract}
 Using the theory of fibre bundles, we provide several equivalent intrinsic
 descriptions for the Hilbert spaces
 of $n$ ``free'' nonrelativistic and relativistic plektons in two
 space dimensions.
 These spaces carry a ray representation of the
 Galilei group and a unitary representation of the Poincar\'{e} group
respectively.
 In the relativistic case we also discuss the situation where the braid group
 is replaced by the ribbon braid group.
\end{abstract}
\section{Introduction}
The last years have seen a rising interest in the theory of particles
in space time dimensions two and three with strange statistics. The possibility
for this was first discovered by Leinaas and Myrheim \cite{LM} , who realized
that the
braid group has to replace the permutation group. Models for particles
with a one-dimensional representation of the braid group where first discussed
by  F.Wilczek \cite{Wi}, who coined the name anyons for particles with these
new statistics (see also \cite{GMS}). In the general case, where the finite
dimensional irreducible
representation of the braid group is not one-dimensional, one speaks of
plektons \cite{FRS}.

Unfortunately so far (non) relativistic fields describing  ``free'' plektons
and having suitable localization properties have not yet been constructed.
However, plektonic structures have been discovered and analyzed within the
context of algebraic quantum field theory and one has an understanding of the
associated Haag - Ruelle scattering theory \cite{FM2,FGR}.

It is the aim of this article to provide a direct intrinsic construction of the
Hilbert spaces of ``free'' nonrelativistic and relativistic plektons in
two space dimensions. These Hilbert spaces will carry a unitary ray
representation of the Galilei group and a unitary representation of the
universal covering group of the
Poincar\'{e} group respectively. We will employ the theory of vector bundles.

In the $n$ particle case the base space will be given as the set of all
momenta of $n$ indistinguishable particles. The universal covering space
will be a principal bundle with the braid group $B_n$ as structure group.
For any finite dimensional unitary
representation of $B_n$ there is an associated
bundle. Given such a representation, by definition the Hilbert space
for $n$ plektons is the space of square integrable sections in this vector
bundle.

Also these bundles will be shown to be homogeneous bundles with respect to the
homogeneous
Galilei and Poincar\'{e} group respectively. This will be the key ingredient
for constructing the unitary (ray) representations.

If one starts with an AFD (=approximately finite dimensional) representation
$\varrho _\infty$ of the infinite braid group $B_\infty$ (see e.g. \cite{We}),
then $\varrho_\infty$ gives rise to a finite dimensional representation
$\varrho_n$ of
$B_n$ for each $n$ and this in turn defines Hilbert spaces ${\cal H}_n$
via the above construction. The resulting Hilbert space
\BEQ \label{1.1}
  {\cal H} = \Bcc \oplus \bigoplus_{n\ge 1} \calH _n
\EEQ
then might serve as a substitute for the bosonic or fermionic Fock spaces.
It remains to construct ``free'' fields which transform correctly under the
Galilei
group and Poincar\'{e} group respectively and which in the relativistic case
have suitable
localization properties. This problem is still open and the difficulties
associated to
this program will also
become apparent within our set-up. In fact, as suggested by the work
of Buchholz and Fredenhagen, these relativistic fields should be localized in
space-like cones \cite{BF}. As is well known in
the ordinary case of spin $\halb $ particles, in the momentum formulation of
the
one particle theory one has to go from a spin basis to a spinor basis as a
necessary step in obtaining anticommuting local spinor fields. At the moment it
is unclear to us what the corresponding procedure should be in two space
dimensions,
where the spin is not quantized. It is known that in case the spin is not
integer or half integer the fields necessarily have to obey braid group
statistics \cite{FM1},\cite{F}.
However, there is the possibility that the
problem may be tackled if instead of the braid group one considers the ribbon
braid group. This corresponds to the situation where each plekton carries an
additional degree of freedom given by a point on the circle.

The article is organized as follows. In section 2 we present the Hilbert space
construction and provide several equivalent formulations. In the case of
anyons the associated bundles are known to be trivial \cite{Do}, but we provide
a new constructive proof.
In the nonrelativistic case these line bundles are also trivial when considered
as homogeneous bundles with respect to the homogeneous Galilei group.
In the relativistic case, however, these bundles are not trivial when viewed as
homogeneous bundles with respect to the Lorentz group, unless the
representation
of $B_n$ is the trivial one.
The proof will be given in appendix B.

In section 3 and 4 we construct the unitary (ray) representations of the
Galilei and Poincar\'{e} groups respectively. In section 5 we give the ribbon
braid construction in the relativistic case and formulate the relativistic
covariance.

This work is based in part on previously unpublished remarks \cite{S}
and a diploma thesis \cite{Mu}.
\paragraph{Acknowledgements:} At various stages of the work the authors have
profited from discussions with K.Fredenhagen, T.Friedrich,
J.Mather, W.M\"{u}ller and F.Nill. D.R.Grigore has kindly pointed out an error
in an earlier version of Appendix A.

\section{The Plektonic Hilbert Spaces}
Let $M$ be a smooth manifold of dimension $\ge 2$ which is connected and
simply connected, i.e. both $\pi_0(M)$ and $\pi_1(M)$ are trivial.
In the $n$ fold product $M^{\times n}$ let  $D_n$ be the set of
points ${\bf m}=(m_1,\ldots ,m_n)\; (m_i \in M)$ with $m_i=m_j$ for at least
one
pair $(m_i,m_j)$ with $i\ne j$. Let $S_n$ be the permutation group of $n$
elements.
$S_n$ obviously acts as a transformation group on $M^{\times n}$ (on the
right),
leaving $D_n$ invariant:\\
\[
 ({\bf m}\pi)_i=m_{\pi(i)} \qquad,\pi \in S_n, {\bf m} \in M^{\times n}.
\]
We introduce the space
 \[ ^n M =(M^{\times n}\setminus D_n)\,/\,S_n. \]
As is well known $\pi_1(^nM)\cong S_n$ such that $M^{\times n}\setminus D_n$
is the universal covering space of $^nM$ except when ${\rm dim } M =2.$
Thus the interesting cases arise when $M\cong \Bb^2$ or $M\cong S^2$ with
$\pi_1(^nM)$ being called the corresponding braid groups for $n$ elements (see
e.g. \cite{Bi}). For the case $M=\Bb^2$ we will write this group as $B_n$,
following standard notation.
\begin{Ex}
 In the case $M=\Bb^2 $ we will view $M$ as the momentum space of a
nonrelativistic
 particle in two space dimensions with points being denoted by $p=(p_1,p_2).$
\end{Ex}
\begin{Ex}\\
 In the case $M=\Hyp=\Big\{p=(p^0,p^1,p^2) \in \Bb^3,\;
p^0=\Big((p^1)^2+(p^2)^2
 +m^2 \Big)^\halb\Big\} \cong \Bb^2,$ we will view $M$ as the energy-momentum
 space (the forward mass shell) of a relativistic free particle of mass $m >
0.$
\end{Ex}
The spaces $^n\Bb^2$ and $^n\Hyp$ will be viewed as the momentum space for $n$
identical nonrelativistic and relativistic free particles, respectively.
Points in these spaces will be denoted by $\bfp.$ Let $^n\tilde{\Bb}^2$
and $^n\tilde{V}^{+,m}$ be their universal covering spaces, with points being
denoted by $\ptild.$ Thus we have the principal $B_n$ bundles
\[
\BA{ccc}
 B_n&\longrightarrow & ^n\tilde{\Bb}^2 \\
    &                & \qquad\Biggm\downarrow\prtild \\
    &                & ^n\Bb^2
\EA\qquad ,\qquad
\BA{ccc}
 B_n&\longrightarrow & ^n\tilde{V}^{+,m} \\
    &                & \qquad\Biggm\downarrow\prtild \\
    &                & ^n\Hyp
\EA
\]
which are of course diffeomorphic.

Let $F$ be a finite dimensional Hilbert space and $\varrho:b\mapsto \varrho(b)
\;(b\in B_n)$ a unitary representation of $B_n$ in $F$ (not necessarily
irreducible).
This defines associated hermitean vector bundles
$^n\tilde{\Bb}^2\times_{\varrho,B_n}F$ and
$^n\tilde{V}^{+,m}\times_{\varrho,B_n}
F.$ In order not to burden the notation, we will simply write $\calFtild$
for these hermitean vector bundles, since $n,F$ and $\varrho$ will be fixed in
what follows and since it will be clear from the context whether we are dealing
with the nonrelativistic or the relativistic case. Similarly, $M$ will from now
on
stand for $\Bb^2$ or $\Hyp.$

By definition, $\calFtild$ is the set of orbits in $^n\tilde{M}\times F$ under
the
following right action of $B_n$ on this space:
$\BA{cr}
 b:(\ptild,f)\mapsto (\ptild\cdot b,\varrho(b^{-1})f)&
  (\ptild \in \nMtild, f\in F,b\in B_n).
\EA $
We denote by
\[ \chitild: \nMtild\times F\to\calFtild  \]
the canonical projection, which by definition is the map associating to
each point $(\ptild,f)$ the orbit on which it lies.

The following well known lemma is the main motivation for our ansatz of a
quantum
mechanical description of particles with braid group statistics.
\begin{Lem}  \label{Lem2.1}
 There is a one-to-one linear correspondence between $C^\infty$ sections
$\xitild$
 in $\calFtild$ and $C^\infty$ functions $\psi$ on $\nMtild$ with values in $F$
which
 obey the  equivariance relation
 \BEQ \label{2.1}
  \psi(\ptild)=\varrho(b)\:\psi(\ptild\cdot b)
 \EEQ
 for all $\ptild \in \nMtild$ and $b\in B_n.$
\end{Lem}
We briefly recall this correspondence. Given $\xitild$, $\:\psi=\psi_{\xitild}$
is
defined as follows. For $\ptild \in \nMtild $ let $\bfp=\prtild (\ptild )
\in {}^nM$
denote the corresponding base point. Then there is a unique $f\in F$ with
$\xitild (\bfp)= \chitild (\ptild ,f)$ and we set $\psi_{\xitild} (\ptild )=f.$
Conversely, given $\psi$, the corresponding $\xitild=\xitild_\psi$ is given by
$\xitild_\psi({\bfp})=\chitild(\ptild,\psi(\ptild))$ for any $\ptild$ with
$\prtild (\ptild )=\bfp.$  These correspondences are easily seen to be well
defined and inverse to each other, and satisfying (\ref{2.1}) .
By going to local trivializations, it follows that
$\xitild_\psi$ is smooth if $\psi$ is and
$\psi_{\xitild}$ is smooth if $\xitild$ is.

Furthermore, let $\langle\;,\;\rangle_\bfp$ be the canonical scalar product on
the fibre in $\calFtild$\\over $\bfp$, i.e.
\BEQ \label{2.2}
 \langle\chitild(\ptild,f),\chitild(\ptild,f')\rangle_{\prtild (\ptild )}
  = \langle f,f'\rangle
\EEQ
where $\langle \; ,\; \rangle $ denotes the scalar product in $F.$ Then we have
\BEQ \label{2.3}
 \left\langle \xitild (\bfp ),\xitild'(\bfp )\right\rangle_\bfp=\left\langle
 \psi_{\xitild}(\ptild ), \psi_{\xitild'}(\ptild )\right\rangle
\EEQ
for all $\ptild $ with $\bfp =\prtild (\ptild ).$
Let $d\mu(\bfp )$ denote the canonical volume form on $^nM$ inherited from
Lebesgue
measure on $\Bb^2$ if $M=\RZ $ and from the Lorentz invariant measure
$d\mu(p)=\halb \Big((p^1)^2+(p^2)^2+m^2\Big)^{-\halb}\:dp^1dp^2$ on $\Hyp$ if
$M=\Hyp$.
By $L^2(\calFtild)$ we denote the Hilbert space completion of the space of
smooth sections of $\calFtild$ having finite norm with respect to the
scalar product
\BEQ \label{2.4}
 \langle \xitild,\xitild'\rangle=\int_{^nM}\left\langle\xitild (\bfp
),\xitild'(\bfp )
  \right\rangle_{\bfp} d\mu(\bfp ).
\EEQ
Similarly, let $L^2_{eq}(\nMtild,F)$ denote the Hilbert space completion of the
space
of smooth functions $\psi$ from $\nMtild$ into $F$ satisfying (\ref{2.1})
and having finite norm with respect to the scalar product
\BEQ   \label{2.5}
 \langle\psi,\psi'\rangle=\int_{^nM}\left\langle\psi(\ptild),\psi'(\ptild)
  \right\rangle d\mu(\bfp).
\EEQ
Note that by (\ref{2.1}) the integrand on the r.h.s. is a function of $\bfp=
\prtild (\ptild )$ only, such that the definition (\ref{2.5}) of the scalar
product
makes sense.

Also by (\ref{2.3}) the above map $\xitild \mapsto \psi_{\xitild}$ extends to a
unitary map
$V$ from $L^2(\calFtild)$ onto $L^2_{eq}(\nMtild,F).$
\begin{Post} \label{Post2.2}
 The quantum mechanical Hilbert space for $n$ ``free'' plektons and for a given
 unitary representation $\varrho$ of $B_n$ on $F$ is given by the space
 $L^2(\calFtild)$ or alternatively by $L^2_{eq}(\nMtild,F).$
\end{Post}
In other words,we view a square integrable section $\xitild\in L^2(\calFtild)$
as the
mathematical formulation of a wave function over the space of $n$
indistinguishable
momenta. Also in terms of the equivalent description by the function
$\psi_{\xitild}=
V\xitild$, condition (\ref{2.1}) replaces the familiar condition where the wave
functions
are required to be symmetric (Bose-Einstein statistics) or antisymmetric
(Fermi-Dirac
statistics). In fact, we may recover these cases as special cases in the
present
context as follows.

Let $\Sbos$ denote the trivial representation of $S_n$ and $\Sfer$ the
alternating
(one-dimensional) representation $\pi\mapsto\sgn(\pi)$ of $S_n$.
Also let $\tau:B_n \to S_n$ be the canonical homomorphism with kernel being the
pure
braid group $PB_n.$
Then we have one-dimensional representations $\rhobos(b)=\Sbos(\tau(b))$ and
$\rhofer=\Sfer(\tau(b))$ of $B_n.$ Functions $\psi$ on $\nMtild$ satisfying
(\ref{2.1}) descend to functions on $M^{\times n}\setminus D_n$, since
$\nMtild$
is a principal $PB_n$ bundle over this space (see below).
The resulting functions $\psi_{\rm bos}$ and $\psi_{\rm fer}$ satisfy the
relations
\BEA \label{2.6}
 \psi_{\rm bos}(\bfp)&=\Sbos(\pi)\:\psi_{\rm bos}(\bfp \pi)&{\rm and }
\nonumber \\
 \psi_{\rm fer}(\bfp)&=\Sfer(\pi)\:\psi_{\rm fer}(\bfp \pi)&
\EEA
for $\bfp \in M^{\times n}\setminus D_n,\ \pi\in S_n.$

We turn to an alternative description of a Hilbert space for $n$ ``free''
plektons.
Recall that $M$ is $\RZ$ or $\Hyp$. Consider the space $M^{\times n}\setminus
D_n$ which is a regular covering space of $^nM$
\[  M^{\times n}\setminus D_n \mathop{\longrightarrow}\limits_\pr {} ^nM.  \]
Its universal covering space with projection $\prttild$ may be identified with
$\nMtild$ such that we have a commutative diagram
\BEQ \label{2.7}
\BA{ccc}
 \nMtild && \\
 {\scriptstyle \prttild}\biggm\downarrow\qquad&
\mathop{\searrow}\limits^{\prtild}\\
 \MnDn & \mathop{\longrightarrow}\limits_\pr & {}^nM
\EA
\EEQ
More precisely, $\nMtild$ is a principal $PB_n$ bundle over $M^{\times
n}\setminus
D_n$, and $M^{\times n}\setminus D_n$ is an $S_n$ bundle over $^nM.$

Given the representation $\varrho$ of $B_n$, we may form the hermitean vector
bundle $\Ztilde{\calF} = \nMtild\times_{\varrho,PB_n}F$ over $\MnDn$, whose
projection
we also denote by $\prttild.$  The map $\Ztilde{\chi}:\nMtild\times
F\to\Ztilde{\calF}$
is defined in analogy to $\chitild$. The map $\Pr:\Ztilde{\calF}\to\calFtild$
is defined by
associating to each $PB_n$ orbit in $\nMtild\times F$ its $B_n$ orbit.
Thus $\Pr$ is a vector bundle map lifting $\pr$:\\
\[ \BA{ccc}
 \Ztilde{\calF} & \mathop{\longrightarrow}\limits^{\Pr} & \calFtild \\
 {\scriptstyle \prttild} \biggm\downarrow \qquad && \qquad\biggm\downarrow
  {\scriptstyle \prtild} \\
 \MnDn& \mathop{\longrightarrow}\limits^\pr & {}^nM
\EA \]\\
Here we have again denoted by $\prttild$ and $\prtild$ the canonical
projections
induced by the projections in the corresponding principal bundles.

Since by construction the fibres in $\Ztilde{\calF}$ are isomorphic to those in
$\calFtild$,
there exists a linear pullback ${\rm Pr}^\ast$ on the space of $C^\infty$
sections
\[ {\rm Pr}^\ast:\Gamma (\calFtild )\to\Gamma (\Ztilde{\calF} ).  \]
By construction, $\frac{1}{n!}{\rm Pr}^\ast$ is isometric with respect to the
scalar
product (\ref{2.5}).
We want to characterize its image. The following result is well known in the
theory of covering spaces. Its relevance in the present context was first
observed by F.Nill, extending a previous remark in \cite{S}.

We claim that $\Ztilde{\calF}$ is a homogeneous $S_n$ bundle, i.e. the action
of $S_n$
on $\MnDn$ lifts to a (right) action on $\Ztilde{\calF}$ which we will write as
\BEQ  \label{2.8}
\begin{array}{ccc}
\Ztilde{\calF} & \mathop{\longrightarrow}\limits^{S(\pi)} &\Ztilde{\calF}\\
{\scriptstyle \prttild}\biggm\downarrow\qquad &&
\qquad\biggm\downarrow{\scriptstyle \prttild} \\
\MnDn & \mathop{\longrightarrow}\limits^\pi & \MnDn
\end{array}
\EEQ
such that $S(\pi)S(\pi')=S(\pi'\pi).$

To see this, let $\{b(\pi)\}_{\pi\in S_n}$ be an arbitrary family of elements
in
$B_n$ with $\tau(b(\pi))=\pi$ for all $\pi\in S_n.$ Given an arbitrary element
$\Ztilde{\chi} (\ptild ,f)$ in the fibre in $\Ztilde{\calF}$ over
$\bfp=\prttild (\ptild) \in
\MnDn$, set
\BEQ \label{2.9}
 S(\pi)\:\Ztilde{\chi}(\ptild,f)=\Ztilde{\chi}\Big(\ptild\cdot
b(\pi),\varrho\big(b(\pi)\big)^{-1}f\Big).
\EEQ
Since $PB_n$ is a normal subgroup of $B_n$, this definition is easily seen to
make sense and to be independent of the particular choice of the family
$\{b(\pi)\}_{\pi\in S_n}.$ Also the diagram
(\ref{2.8})
is commutative, since
obviously $\prttild\:S(\pi )\:\Ztilde{\chi} (\ptild ,f)=\bfp \pi.$
This defines a unitary representation of $S_n$ on $\Gamma(\Ztilde{\calF})$ via
\BEQ
\Big(U(\pi)\:\Ztilde{\xi}\Big)(\bfp)=S(\pi)^{-1}\:\Ztilde{\xi}(\bfp\pi).
\EEQ

Let $\Gamma_{inv}(\Ztilde{\calF})$ be the linear subspace of $\Gamma
(\Ztilde{\calF} )$
consisting of all sections $\Ztilde{\xi}$ satisfying
\BEQ \label{2.10}
 \Ztilde{\xi} =U(\pi)\, \Ztilde{\xi} \:.
\EEQ
This compares with the special situation described in (\ref{2.6}). Define the
linear
operator $P$ on $\Gamma (\Ztilde{\calF} )$ by
\BEQ \label{2.11}
 P\Ztilde{\xi}=\frac{1}{n!}\sum_{\pi\in S_n}U(\pi)\:\Ztilde{\xi}\:.
\EEQ
It is easy to see that $P^2=P$ and that $P^\ast=P$ with respect to the
scalar product in $\Gamma (\Ztilde{\calF} ).$ Furthermore by standard arguments
\BEQ \label{2.12}
 \Gamma_{inv}(\Ztilde{\calF} )=P\:\Gamma (\Ztilde{\calF} ).
\EEQ
We now have the
\begin{Lem} \label{Lem2.4}
 The following equality is valid:
 \BEQ \label{2.13}
  \Gamma_{inv}(\Ztilde{\calF})={\rm Pr}^\ast\:\Gamma (\calFtild ),
 \EEQ
 such that $\frac{1}{n!}\,{\rm Pr}^\ast$ defines an isometry between
$\Gamma(\tilde{\calF})$ and
 $\Gamma_{inv}(\Ztilde{\calF})$.
\end{Lem}
\begin{Proof}
 Given $\Ztilde{\xi}\in \Gamma_{inv}(\Ztilde{\calF})$, define $\xitild\in\Gamma
(\calFtild )$ as
 follows. For $\bfp'\in {}^nM$ choose $\bfp\in\pr^{-1}\{\bfp'\}\subseteq\MnDn$.
 Now write $\Ztilde{\xi} (\bfp )=\Ztilde{\chi} (\ptild ,f)$ for a suitable
 $f\in F$ and $\ptild\in\nMtild$
 with $\prttild (\ptild )=\bfp$ and hence $\bfp'=\prtild (\ptild ).$
 If we set $\xitild (\bfp ')= \chitild (\ptild ,f)$ then by (\ref{2.10}) it is
easy to see
 that $\xitild $ is well defined. Going to a local coordinate system,
 $\xitild$ is also seen to be smooth. Obviously we have $\Ztilde{\xi} = {\rm
Pr}^\ast\:\xitild$.
 Conversely, given $\xitild\in\Gamma (\calFtild )$, $\Ztilde{\xi} ={\rm
Pr}^\ast\:\xitild $ is easily seen
 to satisfy (\ref{2.10}), q.e.d.
\end{Proof}

Let $L^2_{inv}(\Ztilde{\calF} )$ be the closed subspace of $L^2(\Ztilde{\calF}
)$ spanned
by $\Gamma_{inv}(\Ztilde{\calF} )$. Also $P$ extends to an orthogonal
projection
on $L^2(\Ztilde{\calF})$ such that $L^2_{inv}(\Ztilde{\calF})={\rm Range }\:P$.
Furthermore, $\frac{1}{n!}{\rm Pr}^\ast$ extends to a unitary map from
$L^2(\calFtild)$ onto
$L^2_{inv}(\Ztilde{\calF}).$ We collect these results in the third
characterization of
the Hilbert space for $n$ plektons.
\begin{Cor} \label{Cor2.5}
 Via the map $\frac{1}{n!}{\rm Pr}^\ast$ the space $L^2(\calFtild)$ is
unitarily equivalent to the
 linear subspace $L^2_{inv}(\Ztilde{\calF})={\rm Range }P$
 of $L^2(\Ztilde{\calF})$ consisting of
 square integrable sections $\Ztilde{\xi}$ in $\Ztilde{\calF}$ satisfying
(\ref{2.10})
 for almost all $\bfp\in\MnDn.$
\end{Cor}
In the context of algebraic quantum field theory, the space of Haag-Ruelle
scattering
states describing $n$ identical particles has been shown by Fredenhagen,
Gaberdiel and
R\"{u}ger \cite{FGR} to have the same structure
as $L^2(\Ztilde{\calF})$ for a certain class of massive theories.
The anyonic case is also covered in \cite{FM2}.

In case the representation $\varrho$ of $B_n$ is one-dimensional, which is the
anyonic situation, we have the
\begin{Thm} \label{Thm2.6}
 In the anyonic case the line bundles $\calFtild$ are trivial.
\end{Thm}
This observation was first made by J.S.Dowker \cite{Do}, based on Arnol'd's
result
that $H^2({}^nM,{\bf Z})=0$ \cite{Ar} and the classification theorem
of Cartan, Kostant, Souriau and Isham.
It was rediscovered by M.Gaberdiel \cite{Ga} and is implicitly
contained in \cite{Wu} and in \cite[p.564]{FM2}. Here we
provide an alternative, constructive
\begin{Proof}:
The triviality stems from the fact that the first homology group of $^nM$
with integer coefficients is free, that is ${\rm Tor}H_1({}^nM)=0.$
This can be seen as follows.

$H_1({}^nM)$ is the abelianized fundamental group $B_n/[B_n,B_n]$ of $^nM$,
and so the set of unitary one-dimensional representations of $B_n$ is just the
group $\Hom (H_1({}^nM),S^1)$. This group is easily seen to be isomorphic to
\BEQ {\rm Tor}H_1({}^nM)\oplus\frac{\Hom (H_1({}^nM),\Bb)}{\Hom
 (H_1({}^nM),{\bf Z})}
\EEQ
and hence, using the universal coefficient theorem,
we have the isomorphism
\BEQ\label{B1}
 \Hom (H_1({}^nM),S^1)\cong
  {\rm Tor}H_1({}^nM)\oplus\frac{H^1({}^nM,\Bb)}{H^1({}^nM,{\bf Z})}.
\EEQ
The braid group is generated by the elementary braids $b_k$ ($k=1,\ldots,n-1$),
where $b_k$ is the homotopy class of a closed path in ${}^nM$ through a fixed
base
point $\bfp_0$, whose lift to $\MnDn$ interchanges the $k$th and $(k+1)$st
components (see e.g. \cite{Bi}).
They obey the relations
\BEA \label{BraidRel}
 &b_kb_l=b_lb_k &\mbox{ for }k,l\in\{1,\ldots,n-1\}\;\mbox{ and
}|k-l|>1;\nonumber\\
 &b_kb_{k+1}b_k=b_{k+1}b_kb_{k+1}&\mbox{ for }k\in \{1,\ldots,n-1\} .
\EEA
{}From these relations we infer that $H_1({}^nM)$ is freely generated
by the equivalence class $[b_1]$ of any one of the braid group generators $b_1,
\ldots,b_{n-1}$, which are all homology-equivalent.
In particular, ${\rm Tor}H_1({}^nM)=0$, and so (\ref{B1}) implies that
every one-dimensional unitary representation $\varrho$ of the braid group can
be
written as
\BEQ\label{B2}
 \varrho([\beta])=\exp 2\pi i\int_\beta \omega\;,
\EEQ
where $\beta$ is a closed path in ${}^nM$ through the base point $\bfp_0$,
 $[\beta]\in B_n$ its homotopy class and
$\omega$ is a closed 1-form, which is uniquely determined
modulo a closed integer 1-form by the representation $\varrho$.

To be more constructive, consider the 1-form $\frac{1}{\pi}\sum_{k<l}d
\theta^{kl}$ on $\MnDn$, where $\theta^{kl}$ is the angle between the $l$th
and $k$th point in $\RZ.$
This 1-form is the pullback of a unique 1-form
$\omega_1$ on ${}^nM$, whose cohomology class is the dual base to the base
$[b_1]$ of $H_1({}^nM)$, in the sense that $\int_{b_1}\omega_1=1.$
Hence the cohomology class of the 1-form $\omega$ of formula (\ref{B2}) can be
expressed
as $[\omega]=r\cdot[\omega_1]$, where  $r\in \Bb\, {\rm mod }\, {\bf Z}$ is
uniquely determined by the representation $\varrho$. This form was also
implicitly
used in \cite{Wu} and \cite[p.564]{FM2}.

We can exploit formula (\ref{B2}) to construct a nowhere vanishing section
$\xitild$
of the line bundle $\calFtild$. To this end we consider the universal covering
space ${}^n\tilde{M}$ as the set of (fixed end point) homotopy classes of paths
in
${}^nM$ starting from the base point $\bfp_0$. The homotopy class of a
path $\alpha$ will be denoted by ${\rm cls}(\alpha).$ The covering projection
is then given as $\prtild:{\rm cls}(\alpha)\mapsto \alpha(1),$ and the braid
group
acts on ${}^n\tilde{M}$ on the right via $[\beta]:{\rm cls}(\alpha)\mapsto{\rm
cls}
(\alpha)\cdot [\beta] = {\rm cls} (\beta^-\ast\alpha),$ where $\ast$ denotes
the canonical
composition of paths and $\beta^-$ is the inverse path defined by $\beta^-
(t):=\beta(1-t).$  Now we can define a section $\xitild\in\Gamma(\calFtild)$
by setting
\BEQ\label{B3}
 \xitild(\bfp)=\chitild\Big({\rm cls}(\alpha),e^{2\pi
i\int_\alpha\omega}\Big)\;,
\EEQ
where $\omega$ is the 1-form of formula (\ref{B2}) corresponding to the
representation $\varrho$, and
$\alpha$ is any path in ${}^nM$ starting from the base point $\bfp_0$ and
ending
in $\bfp.$ The r.h.s. is easily seen to be independent of the path $\alpha$,
and
so the definition makes sense.
The section $\xitild$ vanishes nowhere and hence trivializes the line bundle
$\calFtild$, q.e.d.
\end{Proof}
\paragraph{Remark :}
The manifold ${}^n\RZ$ has another description which is algebraic geometric
and which is given as
follows. In ${\Bcc}^n$ with points denoted by $(z_1,\ldots,z_n)$
($z_i\in{\Bcc}$) consider the polynomial
\BEQ Q(z_1,\ldots,z_n)=\prod_{i<j} (z_i-z_j)^2 \:. \EEQ
Using the elementary symmetric functions
\BEAast
 \sigma_1&=&z_1+\cdots+z_n\\
 \sigma_2&=&z_1z_2+z_1z_3+\cdots+z_{n-1}z_n\\
 &\cdots&\\
 \sigma_n&=&z_1z_2\cdots z_n\:,
\EEAast
$Q(z_1,\ldots,z_n)$ may be written as a polynomial $\tilde{Q}(\sigma_1,\ldots,
\sigma_n)$ in the $\sigma_i$'s (see e.g. \cite[p.102]{v.d.W}).
$\tilde{Q}(\sigma_1,\ldots,\sigma_n)$ is obviously weighted homogeneous in the
$\sigma_i$'s of type $\left(n(n-1),\halb n(n-1),\ldots,n-1\right)$,
i.e. the relation
\BEQ
 \tilde{Q}\left(e^{\frac{c}{n(n-1)}}\sigma_1,e^{\frac{2c}{n(n-1)}}\sigma_2,
 \ldots,e^{\frac{c}{n-1}}\sigma_n
 \right)=e^c\cdot\tilde{Q}(\sigma_1,\sigma_2,\ldots,\sigma_n)
\EEQ
holds for all complex $c$ (see e.g. \cite[p.75]{Mi}).

Now ${}^n\RZ$ is diffeomorphic to the set $\Big\{(\sigma_1,\ldots,\sigma_n)\in
  {\Bcc}^n \Bigm| \tilde{Q}(\sigma_1,\ldots,\sigma_n)\neq 0\Big\}$, such that
${}^n\RZ$ is diffeomorphic to the complement of a complex hypersurface in
${\Bcc}^n$. Moreover, this set is fibred over the circle via the map
\BEQ (\sigma_1,\ldots,\sigma_n)\mapsto \frac{\tilde{Q}(\sigma_1,\ldots,
 \sigma_n)}{|\tilde{Q}(\sigma_1,\ldots, \sigma_n)|}\:.    \EEQ

As an example, we have e.g. ${}^2\RZ\cong\RZ\times\Bb^+\times T$ (where $T$ is
the unit
circle). In particular ${}^2\RZ$ has the homotopy type of the unit circle on
which all line
bundles are trivial.

This observation might lead to a classification of the possible nontrivial
vector bundles
over this space. Note, however, that the critical points of $\tilde{Q}$ in
general are not
isolated, such that the results in \cite{Mi} concerning the structure of the
fibres
are not applicable.

We also remark that some of the cohomology of ${}^n\RZ$ is known
(see e.g. \cite[p.29]{Ar}, whose results are cited in \cite{Bri}).
In particular, for $i>1$ all of the cohomology groups $H^i({}^nM,{\bf Z})$ are
finite, and hence for every vector bundle over ${}^nM$ there is a $k$ such that
the $k$-fold Whitney sum is trivial.
\section{Galilei Covariant Plektons} \label{gali}
In this section we will show that in a natural way $L^2(\tilde{\cal F})$
for the case $M=\RZ$ carries a unitary representation of a central extension
of the (universal covering of the) Galilei group. Recall that in general
quantum mechanical covariance
under a symmetry group requires only ray representations. Provided the symmetry
group is a connected Lie group and provided certain continuity requirements
are fulfilled, ray representations are equivalent to unitary representations of
a suitable central extension (see e.g. \cite{Ba2} and \cite{Va}).

The Galilei group $\calG_3$ in 3 space-time dimensions
is a Lie group and consists of all quadruples $(t,a,v,R)$\\
$\Big(a,v\in\RZ,t\in\Bb,R\in SO(2)\Big)$ with unit element $(0,0,0,\unity)$
and the multiplication law
\BEQ \label{3.1}
 (t,a,v,R)(t',a',v',R')=(t+t',a+Ra'+t'v,v+Rv',RR').
\EEQ
This group is the semidirect product $G_3\semiprod \Bb^3$ of the homogeneous
Galilei group $G_3$ consisting of elements $(0,0,v,R)$ and the space-time
translation
subgroup $\cong\Bb^3$ consisting of elements $(t,a,0,\unity).$
Its universal covering group $\calGtild_3$ is given by all quadruples of the
form
$(t,a,v,\varphi)\: (a,v\in\RZ,\, v,\varphi\in\Bb)$ with the multiplication law
\BEQ \label{3.2}
(t,a,v,\varphi)(t',a',v',\varphi')=\Big(t+t',a+R(\varphi)a'+t'v,v+R(\varphi)v',
  \varphi+\varphi'\Big)
\EEQ
where $R:\Bb\to SO(2)$ is the standard homomorphism
\BEQ\label{3.3}
 \varphi\mapsto\left(
  \begin{array}{cc} \cos\Fi & \sin\Fi\\-\sin\Fi &\cos\Fi \end{array}
 \right).
\EEQ
Again $\calGtild_3$ is a semidirect product $\tilde{G}_3\semiprod \Bb^3$
where $\Gtild_3$ is the subgroup consisting of all elements of the form
$(0,0,v,\Fi).$ In a natural way $\Gtild_3$ is the universal covering group
of $G_3.$ Now it turns out that the set of central extensions of $\calGtild_3$
forms a three
dimensional manifold. The proof is given in appendix A. This result contrasts
with the higher
dimensional case, where the central extensions form a one parameter family. At
the moment the physical relevance of these central extensions is not clear to
us.
Therefore we will restrict attention to the central extensions which
correspond to those in higher dimensions.

The resulting central extensions $\calGtild_3^m$ of $\calGtild_3$ are
parametrized by a real number
$m$. For fixed $m$, $\calGtild_3^m$ is given as the set of quintuples
$(\theta,t,a,v,\Fi)\;(\theta,t,\Fi\in\Bb,\,a,v\in\RZ)$ and the composition
law
\BEQ \label{3.3a}
 (\theta,t,a,v,\Fi)(\theta',t',a',v',\Fi')=\Big(\theta'',t+t',a+R(\Fi)a'+t'v,
  v+R(\Fi)v',\Fi+\Fi'\Big)
\EEQ
with
\BEQ \label{3.3b} \theta''=\theta+\theta'+\frac{m}{2}\Big(\langle
a,R(\Fi)v'\rangle-
 \langle v,R(\Fi)a'\rangle+t'\langle v,R(\Fi)v'\rangle\Big)\:.
\EEQ
Here $\langle\;,\;\rangle$ denotes the canonical scalar product on $\RZ$.
Again $\Gtild_3$ is a subgroup of $\calGtild_3^m$ and the elements
$(\theta,0,0,
0,\unity)$ form the central subgroup of $\calGtild_3^m.$ In an irreducible
unitary representation the central elements will be represented by $\exp
i\tau\theta$,
and such a representation leads to a ray representation of $\calGtild_3$ with
multiplier
$(-\tau)$ times the multiplier defined by the last summand in (\ref{3.3b})
 (see e.g. \cite[theorem 10.16]{Va}). If $m>0$, a choice we will make,
then this parameter has the physical interpretation of a mass. Given this
$m$, we let $\Gtild_3$ act on $\RZ$ via
\BEQ \label{3.4}
 (v,\Fi):p\mapsto (v,\Fi)\cdot p=R(\Fi)p-mv
\EEQ
This induces in a canonical way an action of $\Gtild_3$ on $(\RZ)^{\times n}$
which leaves $D_n$ invariant and which commutes with the action of $S_n.$
This implies that the action of $\Gtild_3$ descends to an action on $^n\RZ$
which we write symbolically as
\BEQ \label{3.5}
 (v,\Fi):\bfp\mapsto(v,\Fi)\cdot\bfp=R(\Fi)\bfp-m\mbox{\boldmath $v$}.
\EEQ
Now this action of $\Gtild_3$ on $^n\RZ$ lifts to an action on the universal
covering space,
making the principal $B_n$ bundle $\nRtildZ$ a homogeneous $\Gtild_3$-bundle
(see \cite[p.63]{Br}). In other words, if we write the action of $\Gtild_3$ on
$\nRtildZ$ as a left action
\[ (v,\Fi):\ptild\mapsto (v,\Fi)\cdot\ptild  \]
then it commutes with the right action of $B_n$:
\BEQ \label{3.6}
 \Big( (v,\Fi)\cdot\ptild \Big)\cdot b=(v,\Fi)\cdot(\ptild\cdot b).
\EEQ
Furthermore, this induces an action of $\Gtild_3$ on the associated bundle
$\tilde{\cal F}$, again lifting the action on the base $^n\RZ$, by setting
\[ (v,\Fi)\cdot\chitild(\ptild,f)=\chitild\Big((v,\Fi)\cdot\ptild,f\Big)\:.
\]
We need some further notation. The pairing
\[ (v,\bfp)\mapsto\langle v,\bfp\rangle=\langle v,\sum_{i=1}^n p_i
 \rangle \]
from $\RZ\:\times\:(\RZ)^{\times n}\setminus D_n$ into $\Bb$ is invariant under
the
action of $S_n$ and hence descends to a pairing from $\RZ\times {}^n\RZ$ into
$\Bb$ denoted by the same symbol. Similarly the map $\bfp\mapsto\langle \bfp,
\bfp\rangle=\sum_{i=1}^n\langle p_i,p_i\rangle$ from $(\RZ )^{\times
n}\setminus
D_n$ into $\Bb$ descends to a map from $^n\RZ$ into $\Bb$ denoted by the same
symbol.
Physically, this means of course that for given mass the total energy and the
total
momentum of a plektonic configuration $\bfp\in{}^n\RZ$ is well defined.

With these preliminaries we now define a unitary representation of
$\calGtild_3^m$
on $L^2(\tilde{\cal F}).$ We first consider the case where $\varrho$ is
irreducible.
For $\xitild \in L^2(\tilde{\cal F})$ set
\BEQ \label{3.7}
 \Big(U(g)\:\xitild\Big)(\bfp)=e^{ins\Fi}e^{i\left(-n\theta+\langle
a,\bfp\rangle+\frac{mn}{2}
 \langle a,v\rangle+\frac{t}{2m}\langle\bfp,\bfp\rangle  \right)}
 \; (v,\Fi)\cdot\xitild\left((v,\Fi)^{-1}\cdot\bfp\right)\:,
\EEQ
where $g=(\theta,t,a,v,\Fi)$ and where $s$ is an arbitrary real parameter
having
  the physical interpretation of a one-particle spin.

We may finally formulate a condition under which the representation (\ref{3.7})
of
$\calGtild_3^m$ descends to a representation of $\calG^m_3$, the central
extension of $\calG_3$
defined analogously to $\calGtild^m_3.$ In particular, $\calGtild^m_3$ is the
universal covering group of $\calG^m_3.$ The representation descends iff $g=
(0,0,0,0,2\pi)$ is represented by the identity. Now we have
$\prtild\Big((0,2\pi)\cdot
\ptild\Big)=\prtild(\ptild)$ for all $\ptild\in\nRtildZ$. Hence the action of
$(0,2\pi)$ is an element of the deck transformation group, which is, on the
other hand, given
by the right action of the structure group $B_n$. Consequently,
there exists $b\in
B_n$ such that $(0,2\pi)\cdot\ptild=\ptild\cdot b$ for all $\ptild$.
By looking at suitable $\ptild$, it is easily seen that
$b=c_n$ for $n\ge2$, where $c_n=(b_1\cdots b_{n-1})^n$ is the generator of the
center of $B_n$
(see e.g. \cite{Bi}). Hence the representation (\ref{3.7}) descends to a
representation of
$\calG^m_3$ iff $\exp2\pi ins=\varrho(c_n^{-1}).$

If $\varrho$ is not irreducible, then $\varrho$ may be decomposed into
irreducible components
$\varrho=\varrho_1\oplus\cdots\oplus\varrho_k$ on $F=F_1\oplus\cdots \oplus
 F_k.$
Accordingly $\tilde{\cal F}$ decomposes into a Whitney sum $\tilde{\cal
F}=\tilde{\cal F}_1\oplus\cdots
\oplus\tilde{\cal F}_k$ giving the decomposition $L^2(\tilde{\cal
F})=L^2(\tilde{\cal F}_1)\oplus \cdots \oplus
L^2(\tilde{\cal F}_k).$ On each of these subspaces the above construction may
be carried out
with possibly different choices of the total spin in each of the components.
\section{Relativistic Plektons}
We turn to the relativistic case and start by recalling some well known facts
in order to establish notation. Elements of the orthochronous Poincar\'{e}
group $\calP_3^\uparrow$ of the 3-dimensional Minkowski space $\calM_3$ may
be written as $(a,\Lambda)$ with $a\in\calM_3$ and $\Lambda\in\Lor$,
the orthochronous Lorentz group. Group multiplication is given by
$(a,\Lambda)(a',\Lambda')=(a+\Lambda a',\Lambda\Lambda')$ with unit element
$(0,\unity)$
such that $\calP_3^\uparrow$ is the semidirect product of $\calM_3$ and
$\Lor.$ A twofold covering of $\Lor$ is given as the subgroup of $SL(2,\Bcc)$
(conjugate to $SL(2,\Bb)$) consisting of elements of the form
\BEQ \label{4.1}
 \left(\BA{cc} \alpha &\beta\\ \bar{\beta}&\bar{\alpha} \EA\right)\;,
 \alpha\bar{\alpha}-\beta\bar{\beta}=1.
\EEQ
The corresponding Lorentz transformation $\Lambda=\Lambda(\alpha,\beta)\in
L_3^\uparrow$ is given as follows.
For $a=(a^0,a^1,a^2)\in\calM_3$ we set
\BEQ \label{4.2a}
 \utilde{a}=\left(\BA{cc} a^0&a^1-ia^2\\a^1+ia^2& a^0 \EA \right)
\EEQ
and define $\Lambda(\alpha,\beta)\,a$ by
\BEQ\label{4.2b}
 \utilde{\Lambda a}=\left(\BA{cc} \alpha&\beta\\ \bar{\beta}&\bar{\alpha}\EA
 \right)\;\utilde{a}\;\left(
 \BA{cc}\bar{\alpha}&\beta \\ \bar{\beta}&\alpha \EA\right).
\EEQ
In particular for given $p=(p^0,p^1,p^2)\in \Hyp$ the element of the form
\BEQ\label{4.3}
 \Big(2m(p^0+m)\Big)^{-\halb}\;\left(\BA{cc}
  p^0+m&p^1-ip^2 \\ p^1+ip^2 &p^0+m \EA\right)
\EEQ
gives rise to an element in $\Lor$ called a boost and is denoted by
$\Lambda(p).$
One has \\
$\Lambda(p)\,(m,0,0)=p.$
The universal covering group $\tilde{L}_3^\uparrow$ of $\Lor$ can be explicitly
written as the set
\BEQ\label{4.4}
 \Big\{(\gamma,\omega)\Bigm|\gamma\in\Bcc,\,|\gamma|<1,\,\omega\in\Bb\Big\}
\EEQ
with the group multiplication $(\gamma,\omega)(\gamma',\omega')=(\gamma'',
\omega'')$ being given by
\BEA\label{4.5}
 \gamma''&=&(\gamma'+\gamma e^{-i\omega'})(1+\gamma\bar{\gamma'}e^{-i\omega'})
  ^{-1} \\
 \omega''&=&\omega+\omega'+\einsi
\log\Big\{\,(1+\gamma\bar{\gamma'}e^{-i\omega'})
 (1+\bar{\gamma}\gamma'e^{i\omega'})^{-1}\,\Big\}\;. \nonumber
\EEA
Here the logarithm is defined in terms of its power series \cite[p.594]{Ba1}.
The corresponding element in the twofold covering of $\Lor$ described above is
then
given as
\BEQ\label{4.6}
 (1-\gamma\bar{\gamma})^{-\halb}\;\left(\BA{cc}
 e^{i\frac{\omega}{2}}&\gamma e^{i\frac{\omega}{2}} \\
 \bar{\gamma}e^{-i\frac{\omega}{2}} & e^{-i\frac{\omega}{2}} \EA\right).
\EEQ
The resulting element in $\Lor$ will be denoted by $\Lambda(\gamma,\omega).$
For given $p\in\Hyp$, the element $h(p)=\:\Bigm(\gamma=\gamma(p)=
\frac{p^1-ip^2}{p^0+m}\,,\,\omega=0\Bigm)$ in $\Lortild$ is such that $\Lambda
\Big(h(p)\Big)=\Lambda(p).$

The universal covering group $\tilde{\calP}_3^\uparrow$ of $\calP_3^\uparrow$
is now the semidirect product of $\calM_3$ with  $\tilde{L}_3^\uparrow$,
the group multiplication being given by
\BEQ\label{4.7}
 \Big(a,(\gamma,\omega)\Big)\Big(a',(\gamma',\omega')\Big)=
 \Bigm( a+\Lambda(\gamma,\omega)a'\,,\,(\gamma,\omega)(\gamma',\omega')\Bigm).
\EEQ
In contrast to the nonrelativistic case, but in analogy with the higher
dimensional
case,
$\tilde{\calP}_3^\uparrow$ has no nontrivial central extensions.
$\tilde{L}_3^\uparrow$ acts as a transformation group on $(\Hyp)^{\times n}$
via
\[ (\gamma,\omega):\bfp=(p_1,\ldots,p_n)\mapsto\Big(\Lambda(\gamma,\omega)p_1,
 \ldots,\Lambda(\gamma,\omega)p_n\Big) \]
leaving $D_n$ invariant and commuting with the action of $S_n.$
Hence this action of $\tilde{L}_3^\uparrow$ descends to an action on
$^n\Hyp.$ Analogous to the situation in the nonrelativistic case,
 this action lifts to an action on $^n\tilde{V}^{+,m}$
written as $(\gamma,\omega):\ptild\mapsto(\gamma,\omega)\cdot\ptild$
such that $\Big((\gamma,\omega)\cdot\ptild\Big)\cdot b=(\gamma,\omega)\cdot
(\ptild\cdot b),$
i.e. the principal bundle $^n\tilde{V}^{+,m}$ over $^n\Hyp$ is a
homogeneous $\tilde{\Lor}$ bundle.

The pairing
\BEQ\label{4.7a}
 (a,\bfp)\mapsto\langle a,\bfp\rangle=a^0\sum_{j=1}^n
  p_j^0-\sum_{k=1}^2a^k\sum_{j=1}^np_j^k
\EEQ
from $\calM_3\:\times\:(\Hyp)^{\times n}\setminus D_n$ into $\Bb$ descends
to a pairing from $\calM_3\:\times\:{}^n\Hyp$ into $\Bb$, again denoted
by the same symbol.

We denote by $\widetilde{U(1)}=\Bb$ the abelian subgroup of
$\tilde{L}_3^\uparrow$
consisting of elements of the form $(0,\omega).$ For arbitrary $p\in\Hyp$
and $(\gamma,\omega)\in\tilde{L}_3^\uparrow$, the element
\BEQ\label{4.8a}
 t\Big((\gamma,\omega);p\Big)=h(p)^{-1}\;(\gamma,\omega)\;h\Big(\Lambda
 (\gamma,\omega)^{-1}p\Big)
\EEQ
is in $\widetilde{U(1)}$ and hence may be written in the form
\BEQ\label{4.8b}
 t\Big((\gamma,\omega);p\Big)=
 \Big(0,\Omega\big((\gamma,\omega);p\big)\Big).
\EEQ
In fact, by (\ref{4.8a}) and (\ref{4.5})
\BEA\label{4.8c}
\Omega\Big((\gamma,\omega);p\Big)=
 \omega&+\einsi\log\Big\{&\big(1-\gamma(p)\bar{\gamma}e^{-i\omega}\big)\big(1-
 \bar{\gamma}(p)\gamma e^{i\omega}\big)^{-1}\Big\}\nonumber\\
 &+\einsi\log\biggm\{&\Big(1+\frac{\gamma-\gamma(p)e^{-i\omega}}
 {1-\gamma(p)\bar{\gamma}e^{-i\omega}}
 	\bar{\gamma}
  \big(\Lambda(\gamma,\omega)^{-1}p\big)\Big)\cdot \nonumber \\
 && \Big(1+\frac{\bar{\gamma}-\bar{\gamma}(p)e^{i\omega}}
   {1-\bar{\gamma}(p)\gamma e^{i\omega}}
   \:\gamma\big(\Lambda
  (\gamma,\omega)^{-1}p\big)\Big)^{-1}\biggm\}
\EEA
with
\BEA\label{4.8d}
 \lefteqn{\gamma\Big(\Lambda(\gamma,\omega)^{-1}p\Big)=\Big(-2\gamma p^0+
 \gamma^2e^{i\omega}(p^1+ip^2)+e^{-i\omega}(p^1-ip^2)\Big)\cdot} \\
 &&\Bigm(p^0(1+\gamma\bar{\gamma})-\gamma e^{i\omega}(p^1+ip^2)-\bar{\gamma}
 e^{-i\omega}(p^1-ip^2)+m(1-\gamma\bar{\gamma})\Bigm)^{-1}.\nonumber
\EEA
Note that $\Omega\Big((\gamma=0,\omega);p\Big)=\omega$ for all $\omega$ and
$p$.

With $\bfp\in(\Hyp)^{\times n}\setminus D_n$ the element
\BEQ\label{4.9}
 \Omega\Big((\gamma,\omega);\bfp\Big)=\sum_{i=1}^n\Omega\Big((\gamma,\omega);
  p_i\Big)
\EEQ
defines a map $\Omega$ from $\tilde{L}_3^\uparrow\:\times\:(\Hyp)^{\times n}
\setminus D_n$ into $\Bb$ which is invariant under the action of $S_n$
on $(\Hyp)^{\times n}\setminus D_n.$ Hence $\Omega$ descends to a map from
$\tilde{L}_3^\uparrow\:\times\:{}^n\Hyp$ into $\Bb$ denoted by the same symbol.
Again we start with the case where $\varrho$ is irreducible. Fix $s\in\Bb.$
Then
a unitary representation of $\tilde{\calP}_3^\uparrow$ on $L^2(\calFtild)$ is
defined by
\BEQ\label{4.10}
 \Big(U\big(a,(\gamma,\omega)\big)\:\xitild\Big)(\bfp)=e^{i\langle
a,\bfp\rangle
 +is\Omega\big((\gamma,\omega);\bfp\big)}\;
 (\gamma,\omega)\cdot\xitild\Big((\gamma,\omega)^{-1}\bfp\Big)\,.
\EEQ
For the one-particle case, this is the usual irreducible representation of
$\tilde{\calP}_3^\uparrow$ with mass $m>0$ and spin $s$. The extension to the
general case where $\varrho$ need not be irreducible may be treated as in the
nonrelativistic case.

Similarly (when $\varrho$ is irreducible) we may also formulate a necessary and
sufficient condition
that this representation descends to a representation of $\calP_3^\uparrow$.
In fact, the kernel of the covering homomorphism from
$\tilde{\calP}_3^\uparrow$
onto $\calP_3^\uparrow$ is generated by the element $(\gamma=0,\omega=2\pi).$
But then by (\ref{4.8c}),
$\Omega\Big((0,2\pi);\bfp\Big)=n\cdot2\pi$ for all $\bfp\in{}^n\Hyp$.
By the same arguments as in the previous chapter, the equality
\BEQ\label{4.11}
 e^{2\pi ins}=\varrho(c_n^{-1})
\EEQ
is necessary and sufficient to have a representation of $\calP_3^\uparrow$.

We want to point out that Fr\"{o}hlich and Marchetti \cite{FM2} have shown that
for a certain class of massive quantum field theories the space of Haag-Ruelle
scattering states in the anyonic case carries a representation of
$\calP_3^\uparrow$, which we expect to agree with (\ref{4.10}) for the anyonic
case.

As indicated in the introduction, the difficulties in constructing relativistic
``free'' fields on the ``Fock'' space given by (\ref{1.1}) and with suitable
localization
properties become visible in the construction (\ref{4.10}). In fact, within the
present set-up, it is not clear
how to disentangle the factor $\exp is\Omega\Big((\gamma,\omega);\bfp\Big).$
We recall that in four space-time dimensions this is achieved by switching from
the spin basis to the spinor basis.

In the next section we shall propose an alternative set-up which at least
avoids the
aforementioned difficulty.
\section{Relativistic Particles with Ribbon Braid Statistics}
The ribbon braid group arises in the following context (we restrict
ourselves to the relativistic case, see e.g. \cite{Ni}).
Let $T$ be the unit circle in $\Bcc$ and consider the set
$(\Hyp)^{\times n}\:\times\:T^{\times n}.$ Again $S_n$ acts in a canonical
way on this set, leaving the subset $D_n\times T^{\times n}$ invariant.
We set
\BEQ \label{5.1}
 {}^n(\Hyp\times T)\;=\;\Big(\HypnTn\;\setminus \;(D_n\times T^n)\Big)\Bigm/S_n
\EEQ
and we will write points in this set as $(\bfp,\bft)$. Its universal
covering space will be denoted by ${}^n\big(\widetilde{\Hyp\times T}\big)$
with elements $\pttild$ and canonical projection $\prtild$.
Its structure group $RB_n$ is called the ribbon braid group and can be
described as
follows. On ${\bf Z}^n$, the structure group of $T^{\times n}$, $S_n$ and
hence $B_n$
acts as an automorphism  group. Then we have $RB_n=B_n\semiprod {\bf Z}^n$
as a semidirect product.

There are canonical maps from $^n(\Hyp\times T)$ onto $^n\Hyp$ and $T^{\times
n}
/S_n$
and the images of $(\bfp,\bft)$ will be denoted by $\bfp$ and $\bft$
respectively.

We may now proceed as before. Thus let $\varrho$ be a finite dimensional
unitary representation of
$RB_n$ in a Hilbert space $F$ defining an associated bundle $\calFtild.$
$L^2(\calFtild)$ is
now the Hilbert space for $n$ relativistic plektons with ribbon braid group
statistics. Alternatively we may speak of framed plektons. In defining
$L^2(\calFtild)$ we of course make use of the canonical measure
$d\mu(\bfp,\bft)$
on $^n(\Hyp\times T)$ induced by the Lorentz invariant measure $d\mu(p)$ on
$\Hyp$
and the Haar measure $d\nu(t)$ on $T$.

In the anyonic case, i.e. when $\varrho$ is a one dimensional representation,
the resulting line bundle is again trivial. This may be shown by the methods
used in section 2.

The main observation we need for describing relativistic invariance is that
$\tilde{\Lor}$ acts on $T$ as a left transformation group via
\BEQ\label{5.2}
 (\gamma,\omega)\cdot t=e^{-i\omega}\:\frac{t-\gamma e^{i\omega}}
 {1-t\bar{\gamma}e^{-i\omega}}.
\EEQ
This defines an action of $\tilde{\Lor}$ on $\HypnTn$ leaving $D_n\times
T^{\times n}$ invariant
and commuting with the action of $S_n.$ Hence this action of $\tilde{\Lor}$
descends to an action on $^n(\Hyp \times T)$ written as $(\gamma,\omega):
(\bfp,\bft)\mapsto (\gamma,\omega)\cdot (\bfp,\bft).$
Again, $\tilde{\Lor}$ lifts to an action on the principal $RB_n$ bundle
$^n(\widetilde{\Hyp\times T})$ written as $(\gamma,\omega):\pttild\mapsto
(\gamma,\omega)\cdot\pttild$ such that
\[ \Big((\gamma,\omega)\cdot\pttild\Big)\cdot b=(\gamma,\omega)\cdot
 \Big(\pttild\cdot b\Big)    \] for all $b\in RB_n.$

To construct unitary representations of $\tilde{\calP}_3^\uparrow$
we will take recourse to a certain class of unitary representations of
$\tilde{\Lor}$
in $L^2(T)$, called the principal series. They were first suggested by Bargmann
\cite{Ba1} and then analyzed by Pukanszky \cite{Pu}. These representations are
parametrized
by the pair $(h,\sigma) \quad\Big(-\halb<h\le\halb,\;\sigma \in i\Bb$ (pure
imaginary)$\Big)$.
Unless $h=\halb$ and $\sigma =0$, these representations are irreducible.
The principal series of $SL(2,\Bb)$ are obtained by setting $h=0$ or $h=\halb$.
With a slight modification of the notation in \cite{Sa}, these representations
have the form
\BEQ\label{5.3a}
 \Big( U(\gamma,\omega)\:f\Big)(t)=\tau\Big((\gamma,\omega);t\Big)\,
  f\Big((\gamma,\omega)^{-1}\cdot t\Big)
\EEQ
with $\tau=\tau(h,\sigma)$ given as
\BEQ\label{5.3b}
 \tau\Big((\gamma,\omega);t\Big)=e^{-i\omega h}\:\left(\frac{1+t\bar{\gamma}}
 {1+t^{-1}\gamma}\right)^h \:\big|1+t\bar{\gamma}\big|^{-1-2\sigma}\:
 \big(1+|\gamma|^2\big)^{\halb+\sigma}.
\EEQ
We start by constructing a unitary representation of $\tilde{\calP}_3^\uparrow$
in the one particle case, thus motivating our ansatz.
By definition, the Hilbert space is then given as $L^2\Big(\Hyp\times
T,\:d\mu(p)
d\nu(t)\Big)$. For $\psi$ in this space we set
\BEQ\label{5.4}
 \Big( U\big(a,(\gamma,\omega)\big)\:\psi\Big)(p,t)=e^{i\langle p,a\rangle}\:
\tau\Big((\gamma,\omega);t\Big)
\:\psi\Big(\Lambda(\gamma,\omega)^{-1}p,(\gamma,\omega)^{-1}t\Big)
\EEQ
giving rise to a unitary representation of $\tilde{\calP}_3^\uparrow$. In order
to see how this representation is related to the irreducible unitary
representation
given in the previous section for the one-particle case, we decompose the
representation (\ref{5.4})
as follows. Let $W=W(h,\sigma)$ be the unitary operator defined by
\BEQ\label{5.5}
 \big(W\psi\big)(p,t)=\tau\Big(\big(\gamma(p),0\big);t\Big)\:\psi
 \Big(p,\big(\gamma(p),0\big)^{-1}t\Big)
\EEQ
and set
\BEQ\label{5.6}
 \hat{U}\Big(a,(\gamma,\omega)\Big)=W^{-1}\:U\Big(a,(\gamma,\omega)\Big)\:W.
\EEQ
Then a short calculation gives
\BEQ\label{5.7}
 \Big(\hat{U}\big(a,(\gamma,\omega)\big)\:\psi\Big)(p,t)=e^{i\langle a,p\rangle
 -i\Omega\big((\gamma,\omega);p\big)h}\:
 \psi\Big(\Lambda(\gamma,\omega)^{-1}p,te^{i\Omega
  \big((\gamma,\omega);p\big)}\Big).
\EEQ
For $k\in {\bf Z}$ let $L_k\subset L^2\Big(\Hyp\times T,d\mu(p)d\nu(t)\Big)$
be the image of $L^2\big(\Hyp,d\mu(p)\big)$ under the linear isometric map
\BEQ\label{5.8}
 \psi(p)\mapsto\psi(p,t)=\psi(p)t^k.
\EEQ
We have the direct sum decomposition
\BEQ\label{5.9}
 L^2\big(\Hyp\times T,d\mu(p)d\nu(t)\big)=\bigoplus_{k\in{\bf Z }}L_k.
\EEQ
The above formula shows that $L_k$ is invariant under $\hat{U}\big(a,(\gamma,
\omega)\big).$
More precisely,
\BEA\label{5.10}
 \Big(\hat{U}\big(a,(\gamma,\omega)\big)\:\psi\Big)(p,t)=
 e^{i\langle a,p \rangle+i(k-h)\Omega\big((\gamma,\omega);p\big)}
 \:\psi\Big(\Lambda(\gamma,\omega)^{-1}p,t\Big)
\EEA
for $\psi \in L_k.$ By comparison with (\ref{4.10}) we see that we have an
irreducible
representation of $\tilde{\calP}_3^\uparrow$ on $L_k$ of spin $s=k-h.$
In the language of physicists by analogy to the higher dimensional case, one
may say that $W$ provides the transition from the spinor basis to the spin
basis.

We now generalize this construction to the $n$ particle sector as follows.
For $\bft=(t_1,\ldots,t_n)\in T^{\times n}$ define
\BEQ\label{5.11}
\tau\Big((\gamma,\omega);\bft\Big)=
\prod_{j=1}^n\tau\Big((\gamma,\omega);t_j\Big).
\EEQ
This quantity is invariant under the action of $S_n$ on $T^{\times n}$ and
hence
descends to a function on $T^{\times n}\bigm/S_n$ denoted by the same symbol.
A unitary representation of $\tilde{\calP}_3^\uparrow$ on $L^2(\calFtild)$ is
now
given by
\BEQ\label{5.12}
 \Big(U\big(a,(\gamma,\omega)\big)\:\xitild\Big)\big((\bfp,\bft)\big)=
  e^{i\langle a,\bfp\rangle}\:\tau\big((\gamma,\omega);\bft\big)\;
  (\gamma,\omega)\cdot\xitild\Big((\gamma,\omega)^{-1}(\bfp,\bft)\Big)\:.
\EEQ
\paragraph{Appendix}
\appendix
\section{The Central Extensions of the Galilei Group in 3 Space-Time
 Dimensions}
Here we want to prove our claim that the set of (equivalence classes
of)  central extensions of $\tilde{\calG}_3$, and hence of its Lie algebra,
forms a three dimensional manifold.

As is well known, the set of central extensions of a Lie algebra ${\fraka}$ is
in one-to-one correspondence with its second
cohomology space $H^2({\fraka})$, which is defined as follows.
For $q\in{\bf N}$, let $\Lambda^q(\fraka^\ast)$ denote the linear space
of real valued antisymmetric $q$-linear forms on $\fraka^{\times q}$.
We have the
coboundary operators, defined by
\BEAast
 \delta_1:\quad\;\;\,\fraka^\ast\to\Lambda^2(\fraka^\ast),&
  (\delta_1\lambda)(x,y)\quad=&\lambda([x,y])  \quad\mbox{ and }\\
 \delta_2:\Lambda^2(\fraka^\ast)\to\Lambda^3(\fraka^\ast),&
  (\delta_2\Xi)(x,y,z)=&
  \Xi([x,y],z)+\Xi([y,z],x)+\Xi([z,x],y)
\EEAast
for all $x,y$ and $z\in\fraka.$ The kernel of $\delta_2$ is denoted by
$Z^2(\fraka)$, and the image of $\delta_1$ by $B^2(\fraka).$
The second cohomology space of $\fraka$ is then defined as $H^2(\fraka):=
Z^2(\fraka)/B^2(\fraka)$.

If we denote the generators of rotations, space translations, pure
Galilei trans\-for\-ma\-tions
and time translations
by $l,p_i,n_i$ and $e$ respectively ($i=1,2$), the multiplication
law (\ref{3.2}) of $\tilde{\calG}_3$ implies the following  structure of its
Lie algebra, which will be denoted by $\frakg$ in the sequel:
\BEA
 &[l,p_1]=-p_2,  &[l,p_2]=p_1,\label{app2.1}\\
 &[l,n_1]=-n_2,  &[l,n_2]=n_1,\label{app2.2}\\
 &[n_i,e]=p_i &(i=1,2),\label{app2.3}
\EEA
and all other commutators vanish.

{}From these relations we conclude that an antisymmetric bilinear form $\Xi$ in
$\Lambda^2(\frakg^\ast)$
satisfies the cocycle condition $\delta_2\Xi=0$ iff all of the following
equalities hold:
\BEA
 \Xi(l,p_1)=-\Xi(n_2,e)&,&\Xi(l,p_2)=\Xi(n_1,e)\;,  \label{A.4}\\
 \Xi(p_1,p_2)=0&,&\label{A.5}\\
  \Xi(p_1,n_2)=0=\Xi(p_2,n_1)&,&\Xi(p_1,n_1)=\Xi(p_2,n_2)\;, \label{A.6}\\
 \mbox{ and }\:\Xi(p_i,e)=0\:\:(i=1,2)&.&\label{A.7}
\EEA

The natural grading of the Lie algebra $\frakg$, which is given by the
subalgebras
\[ \frakg_1:={\rm span}\{l\},\;\frakg_2:={\rm span}\{p_1,p_2\},\;
\frakg_3:={\rm span}
\{n_1,n_2\}\mbox{ and }\frakg_4:={\rm span}\{e\}, \]
exhibits $\Lambda^2(\frakg^\ast)$ as a direct sum
\[ \Lambda^2(\frakg^\ast)=\bigoplus_{i=2}^3\Lambda^2(\frakg_i^\ast)\oplus
\bigoplus_{\stackrel{\scriptstyle i,j=1}{i<j}}^4
(\frakg_i\otimes\frakg_j)^\ast,\]
where $(\frakg_i\otimes\frakg_j)^\ast$ denotes the space of bilinear
forms on $\frakg_i\times\frakg_j$. Using this decomposition, we examine the
restrictions $\Xi_{i,j}$
of an arbitrary
cocycle $\Xi\in Z^2(\frakg)$ to the various subspaces $\frakg_i\times\frakg_j$
and note that if $\Xi_{i,j}$ is in $B^2(\frakg)$, we can set it equal to zero
without changing the equivalence class of $\Xi$ in $H^2(\frakg).$

The restrictions of $\Xi$ to $\frakg_2\times\frakg_2$ and
$\frakg_2\times\frakg_4$ are zero due to equations (\ref{A.5}) and (\ref{A.7}),
respectively.

$\Xi_{1,2}$ is the (restriction of the) coboundary of
$\lambda\in\frakg^\ast$ defined by
$\lambda(l)=0$, $\lambda(p_1)=\Xi(l,p_2)$ and $\lambda(p_2)=-\Xi(l,p_1).$

Equation (\ref{A.4}) shows that $\Xi_{3,4}=\delta_1\lambda$ on $\frakg_3\times
\frakg_4$ for the same $\lambda.$

$\Xi_{1,3}$ is seen to be a coboundary, i.e. in $B^2(\frakg)$,
with a similar argument.

$\Xi_{3,3}$ and $\Xi_{1,4}$ are multiples of the cocycles
\BEA
 \Xi^{(1)}(n_1,n_2)=-\Xi^{(1)}(n_2,n_1)&=&1 \;\mbox{ and}\label{app4.1}\\
 \Xi^{(2)}(l,e)&=&1\;,\mbox{ respectively.}\label{app4.2}
\EEA
Finally, with respect to the three linearly independent equations (\ref{A.6}),
$\Xi_{2,3}$ is determined by the cocycle
\BEA
 \Xi^{(3)}(p_1,n_1)=\Xi^{(3)}(p_2,n_2)=1&,&\Xi^{(3)}(p_1,n_2)=0=\Xi^{(3)}
 (p_2,n_1)\;.
  \label{app4.3}
\EEA

None of the cocycles $\Xi^{(1)},$ $\Xi^{(2)}$ and $\Xi^{(3)}$ is a coboundary,
since their respective arguments commute. Similarly, we see by inspection that
the only linear combination which is a coboundary, is the trivial one.

Summing up, $H^2(\frakg)$ is three-dimensional, spanned by the equivalence
classes of the bilinear forms defined in (\ref{app4.1}) to (\ref{app4.3}).

The corresponding central extensions of $\tilde{\calG}_3$ can be found
following
\cite[p.127]{Va}:
Every element $\Xi\in H^2(\frakg)$ determines a multiplier $\omega$ on
$\tilde{\calG}_3
\times\tilde{\calG}_3$
via
\[ \Xi(x,y)=\frac{\partial^2}{\partial s\partial t}\Big(\omega(\exp sx,\exp ty)
 -\omega(\exp sy,\exp tx)\Big)\Big|_{s=t=0} \]
for all $x,y$ in $\frakg$, which in turn defines a central extension
$\tilde{\calG}_3^\omega$
with the multiplication law
\[ (\theta,g)(\theta',g')=\Big(\theta+\theta'+\omega(g,g'),gg'\Big)\;
 (\theta,\theta'\in \Bb, \:g,g'\in \tilde{\calG}_3). \]
$\omega$ and $\tilde{\calG}_3^\omega$ are determined by $\Xi$ uniquely modulo
the relevant equivalence relations. Using the above formula, we find the
following multipliers
$\omega_1$, $\omega_2$ and $\omega_3$ corresponding to $\Xi^{(1)}$, $\Xi^{(2)}$
and $\Xi^{(3)}$
respectively, where $\sigma(\cdot,\cdot)$ denotes the standard symplectic form
on $\RZ$:
\BEAast
 \omega_1(g,g')&=&1/2\;\sigma\Big(v,R(\Fi)v'\Big)\:,\\
 \omega_2(g,g')&=&1/2\; t\Fi'\:\mbox{ and}\\
 \omega_3(g,g')&=&1/2\;\Big(\langle a,R(\Fi)v' \rangle-\langle
v,R(\Fi)a'\rangle
  +t'\langle v,R(\Fi)v'\rangle\Big)
\EEAast
for any $g=(t,a,v,\Fi)$ and $g'=(t',a',v',\Fi')$ in $\tilde{\calG}_3$.
Obviously, the
multiplier $m\cdot\omega_3$ $(m>0)$ corresponds to the central extension
$\tilde{\calG}
_3^m$ considered in section~\ref{gali}.
The possible physical relevance of an
arbitrary central extension has not been clarified yet. We note  that
D.R.Grigore  has determined the corresponding projective unitary
irreducible representations of $\calG_3$ in a recent paper\cite{Grig}.

\section{The Anyonic Line Bundles as (Non) Trivial G-Bundles}

In this appendix we will show that the anyonic line bundles
$\calFtild$
in the nonrelativistic case are also trivial when considered
as $\tilde{G}_3$-bundles (where $G_3$ is the homogeneous Galilei group.
In the relativistic case, however, they are not trivial when viewed as
$\tilde{L}_3
^{\uparrow}$-bundles ($\Lor$ denoting the Lorentz group), unless the
representation
$\varrho$ of $B_n$ defining the line bundle is trivial.

We recall that a $G$-bundle ${\cal E}$ over a $G$-manifold $M$ is defined to be
trivial
iff ${\cal E}$ is diffeomorphic to $M\times E$, where $E$ is a vector space
and the action of $G$ on ${\cal E}$ corresponds to a product action
$g:(m,e)\mapsto(g\cdot m,D(g)e)$, where $g\mapsto D(g)$ is a representation of
$G$ on $E$.
Also $D(g)$ is supposed to be smooth in $g$, if $G$ is a Lie group and if the
action of $G$ on
${\cal E}$ is smooth.

Any equivariant function $F$ on $\nMtild$ with values in $\Bcc^{\times}$
defines a trivialization of the
associated anyonic line bundle $\calFtild$ via the map
$\chitild(\ptild,c)\mapsto\left(
 \prtild(\ptild),F(\ptild)^{-1}c\right)$, and the action of $G$ on $\calFtild$
corresponds to the following action on ${}^nM\times\Bcc$ :
\BEQ \label{appB1}
 g:(\bfp,c)\mapsto \left(g\cdot\bfp,F(\ptild)F(g\cdot\ptild)^{-1}c\right),
\EEQ
where $\ptild\in\nMtild$ is any point in the fibre over $\bfp$.
$F$ trivializes $\calFtild$ as a $G$-bundle iff for all $g\in G$
the function $F(\ptild)\,F(g\cdot\ptild)^{-1}$, which only depends on $\bfp=
\prtild(\ptild)$, is actually independent of $\bfp$. In that case,
\BEQ \label{appB2}
 D(g)=F(\ptild)\,F(g\cdot\ptild)^{-1}
\EEQ
defines a representation of $G$.
Identifying, as done in section 2, points in $\nMtild$ with homotopy classes of
paths in ${}^nM$ starting at the base point $\bfp_0$, the action of $G$ on
$\nMtild$ is given as follows.
Let $\ptild={\rm cls}(\alpha)\in \nMtild$, and let $\gamma$ be any path in $G$
starting at the identity and ending at $g\in G$. Then $g\cdot\ptild$ is the
homotopy class
of the path $(\gamma\cdot\alpha)(t):= \gamma(t)\cdot\alpha(t)$, or,
equivalently,
of the path $\alpha\ast(\gamma\cdot i_{\bfp})$, where $i_{\bfp}$ is the
constant path at
$\bfp=\alpha(1)$.

\subsection{The Nonrelativistic Case}
Here $G=\Gtild_2$, elements of which are denoted by $(v,\Fi)$, and
$M=\RZ$, which will be identified with $\Bcc$. We claim that
in this case the equivariant function
\BEQ \label{appB3}
 F_r\left({\rm cls}(\alpha)\right)=
  \exp\left(2ir\sum_{k<l}\int_{\bar{\alpha}} d\theta^{kl}\right)
  \qquad(\mbox{where }\pr\,\circ \,\bar{\alpha}=\alpha)
\EEQ
from equation (\ref{B3})
satisfies (\ref{appB2}) with $D(v,\Fi)=e^{-irn(n-1)\Fi}$.
To see this, we have to write down explicitly the action of an element $(v,\Fi)
\in\Gtild_2$ on $\ptild\in\nMtild$.
We set $\ptild={\rm cls}(\alpha)$ with
$\alpha(t)=\pr\left(\alpha_1(t),\ldots,\alpha_n(t)\right)$ and take
$\gamma(t)=(tv,t
 \Fi)$ as a path in $G$ starting at the identity  and ending at $(v,\Fi)$. Then
the above definition of the action yields
\BEQ
 (v,\Fi)\cdot{\rm cls}(\alpha)={\rm cls}(\gamma\cdot\alpha)\mbox{ with }
 (\gamma\cdot\alpha)(t)
 =\pr\left(e^{it\Fi}\alpha_1(t)-tmv,\ldots,e^{it\Fi}\alpha_n(t)-tmv
 \right).
\EEQ
Denoting by $\alpha_{kl}(t)$ the path
$\alpha_l(t)-\alpha_k(t)$, we obtain
\[ \int_{\bar{\alpha}} d\theta^{kl}=\int_{\alpha_{kl}}d\theta={\rm
Im}\int_{\alpha_{kl}}
 \frac{dz}{z}={\rm Im}\int_0^1\frac{\dot{\alpha_{kl}}(t)}{\alpha_{kl}(t)}dt, \]
and hence
\[ \int_{\overline{\gamma\cdot\alpha}}d\theta^{kl}=\Fi+\int_{\bar{\alpha}}
 d\theta^{kl}. \]
Inserting these formulas into (\ref{appB3}), we see that
for any $\ptild\in\nMtild$, the desired relation
$F_r(\ptild)\,F_r\left((v,\Fi)\cdot\ptild\right)^{-1}=e^{-irn(n-1)\Fi}$
follows.


\subsection{The Relativistic Case}

Here $G=\Lortild$, which is the covering group of a simple group, and therefore
the one dimensional representation $g\mapsto D(g)$ of equation (\ref{appB2})
is necessarily trivial. Hence the line bundle $\calFtild$ is trivial as an
$\Lortild$-bundle if and only if there is a smooth nowhere vanishing function
$F$ from $\nMtild$ into the complex numbers satisfying
\[
\BA{lll}
 F(\ptild\cdot b)=\varrho(b)^{-1}\,F(\ptild)&\mbox{for all }\ptild\in \nMtild,
  b\in B_n&\mbox{, and}\\
 F(g\cdot\ptild)=F(\ptild)&\mbox{for all }\ptild\in \nMtild, g\in\Lortild&
 .
\EA  \]
A necessary condition for such an $F$ to exist is that the representation
$\varrho$
maps any braid $b\in B_n$, for which there are $\ptild\in\nMtild$ and
$g\in\Lortild$ with
$\ptild\cdot b=g\cdot \ptild$, to the identity.
Now we claim that the generator $b_1$ of the braid group can be written as the
product of two braids $b_{(1)}$ and $b_{(2)}$, each of which satisfies the
above
condition, i.e. there are $\ptild_1,\ptild_2\in\nMtild$ and
$g_1,g_2\in\Lortild$ such that
 $\ptild_j\cdot b_{(j)}=g_j\cdot\ptild_j$ $(j=1,2)$.
Consequently, $\varrho$ has to act trivial on $b_{(1)}b_{(2)}=b_1$.
Since, as a consequence of the defining relations (\ref{BraidRel}), each $b_k$
is conjugate to $b_1$, $\varrho$ is trivial on the generators and hence
on the whole group $B_n$.

To prove our claim, we have to find two points $\bfp_1,\bfp_2\in {}^nM$,
paths $\alpha_1$ and $\alpha_2$ from the base point $\bfp_0$ to $\bfp_1$ and
$\bfp_2$ respectively, $g_1$ and $g_2\in \Lortild$, and $b_{(1)},b_{(2)}\in
B_n$
$=\pi_1({}^nM,\bfp_0)$ such that  $b_{(1)}b_{(2)}=b_1$ and
\BEQ \label{appB5}
 {\rm cls}(\beta_{(j)}\ast\alpha_j)={\rm cls}\left(\alpha_j\ast(\gamma_j
  \cdot i_{\bfp_j})\right) \qquad (j=1,2).
\EEQ
Here $\beta_{(j)}$ is a loop at the base point $\bfp_0$ whose homotopy class is
$b_{(j)}$, and $\gamma_j$ is a path in $\Lortild$ from the
identity to $g_j$.

Taking
$\bfp_1:=\pr\:\left(1,0,e^{i\theta_1},
e^{i2\theta_1},\ldots,e^{i(n-2)\theta_1}\right)$
with $\theta_1=\frac{2\pi}{n-1}$, and $\gamma_1(t):=t\theta_1$,
we get a path $\gamma_1\cdot i_{\bfp_1}$ which is free homotopic as a closed
path in ${}^nM$ to
the following loop $\beta_{(1)}$ at the base point $\bfp_0$:
\begin{center}
Fig. 1
\end{center}
This is equivalent to the statement that there is a path $\alpha_1$ from
$\bfp_0$ to $\bfp_1$
satisfying (\ref{appB5}) (for $j=1$).

Let
$\bfp_2:=\pr\:
\left(1,e^{i\theta_2},e^{i2\theta_2},\ldots,e^{i(n-1)\theta_2}\right)$
with $\theta_2=\frac{2\pi}{n}$, and let $\gamma_2(t):=-t\theta_2$. Then the
closed path
$\gamma_2\cdot i_{\bfp_2}$ is free homotopic to the loop $\beta_{(2)}$ given by
\begin{center}
Fig. 2
\end{center}
Hence there is a path $\alpha_2$ from $\bfp_0$ to $\bfp_2$ satisfying
(\ref{appB5})
for $j=2$, and it remains to be shown that the loop
$\beta_{(1)}\ast\beta_{(2)}$
represents the generator $b_1$ of the braid group. Pictorially
this can be checked as follows:
\begin{center}
Fig. 3
\end{center}
which corresponds to the relations
\BEAast
 b_{(1)}=[\beta_{(1)}]&=&b_1b_{n-1}b_{n-2}\cdots b_2b_1,\\
 b_{(2)}=[\beta_{(2)}]&=&b_1^{-1}\cdots b_{n-1}^{-1}, \mbox{ and hence}\\
 b_{(1)}b_{(2)}&=&b_1.
\EEAast
This proves the claim.

\end{document}